%% file: main.tex
\documentclass[12pt]{article}
\usepackage[colorlinks=true, urlcolor=dullpurple, citecolor=dullred, linkcolor=darkgreen]{hyperref}
\usepackage{mathptmx}
\usepackage{float}
\usepackage{endnotes}
\usepackage{geometry}
\usepackage{rotating}
\usepackage{aas_macros}
\usepackage{comment}
\usepackage[utf8]{inputenc}
\usepackage[english]{babel}
\usepackage{enumitem,amssymb}
\usepackage{ragged2e}
\newlist{thematic}{itemize}{8}
\setlist[thematic]{label=$\square$}
\usepackage{pifont}
\usepackage{cite}
\usepackage{soul}
\usepackage{lscape}
\usepackage{multicol}
\usepackage{multirow}
\usepackage{array}
\usepackage{wrapfig}
\usepackage{graphicx}
\usepackage[range-units=brackets, range-phrase=-]{siunitx}
\usepackage{xcolor}
\definecolor{dullpurple}{rgb}{0.431,0.188,0.534}
\definecolor{darkgreen}{rgb}{0.075,0.302,0.047}
\definecolor{dullred}{rgb}{0.706,0.208,0.192}
\usepackage[babel=true]{microtype}

\newlength{\enumindent}
\setlist{nolistsep}

\usepackage[font=small]{caption}
\newcolumntype{C}[1]{>{\centering\let\newline\\\arraybackslash\hspace{0pt}}m{#1}}

\DeclareSIUnit{\parsec}{pc}
\DeclareSIUnit{\Mpc}{\mega\parsec}
\DeclareSIUnit{\h}{\mathit{h}}
\DeclareSIUnit{\hPerMpc}{\h\per\Mpc}
\newcommand{\hMpc}{h {\rm Mpc}^{-1}}

\newcommand{\minisection}[1]{\noindent \textbf{#1}}
\makeatletter
\renewcommand\section{\@startsection{section}{1}{\z@}%
	{-1.5ex \@plus -0.4ex \@minus -.2ex}%
	{0.8ex \@plus.2ex \@minus .2ex}%
	{\normalfont\Large\bfseries}}
\renewcommand\subsection{\@startsection{subsection}{2}{\z@}%
	{-0.7ex\@plus -0.3ex \@minus -.2ex}%
	{0.4ex \@plus .1ex \@minus .1ex}%
	{\normalfont\large\bfseries}}
\makeatother

\newcommand{\stagetwo}{Stage~{\sc ii}}
\newcommand{\stageone}{Stage~{\sc i}}
\newcommand{\fnl}{f_{\rm NL}}

\input{institutionAliases.tex}

\begin{document}

\pagenumbering{gobble}
{
\raggedright
  
\Large
\noindent Packed Ultra-wideband Mapping Array (PUMA\footnote{\url{https://www.puma.bnl.gov}}):\\
A Radio Telescope for Cosmology and Transients
\linebreak
\normalsize

\minisection{Thematic Areas:}  
Ground Based Project\linebreak
  
\textbf{Primary Contact:}

Name: An\v{z}e Slosar
 \linebreak						
Institution: Brookhaven National Laboratory
 \linebreak
Email: \texttt{anze@bnl.gov}
 \linebreak
Phone:  631-344-8012
 \linebreak

\includegraphics[width=\linewidth]{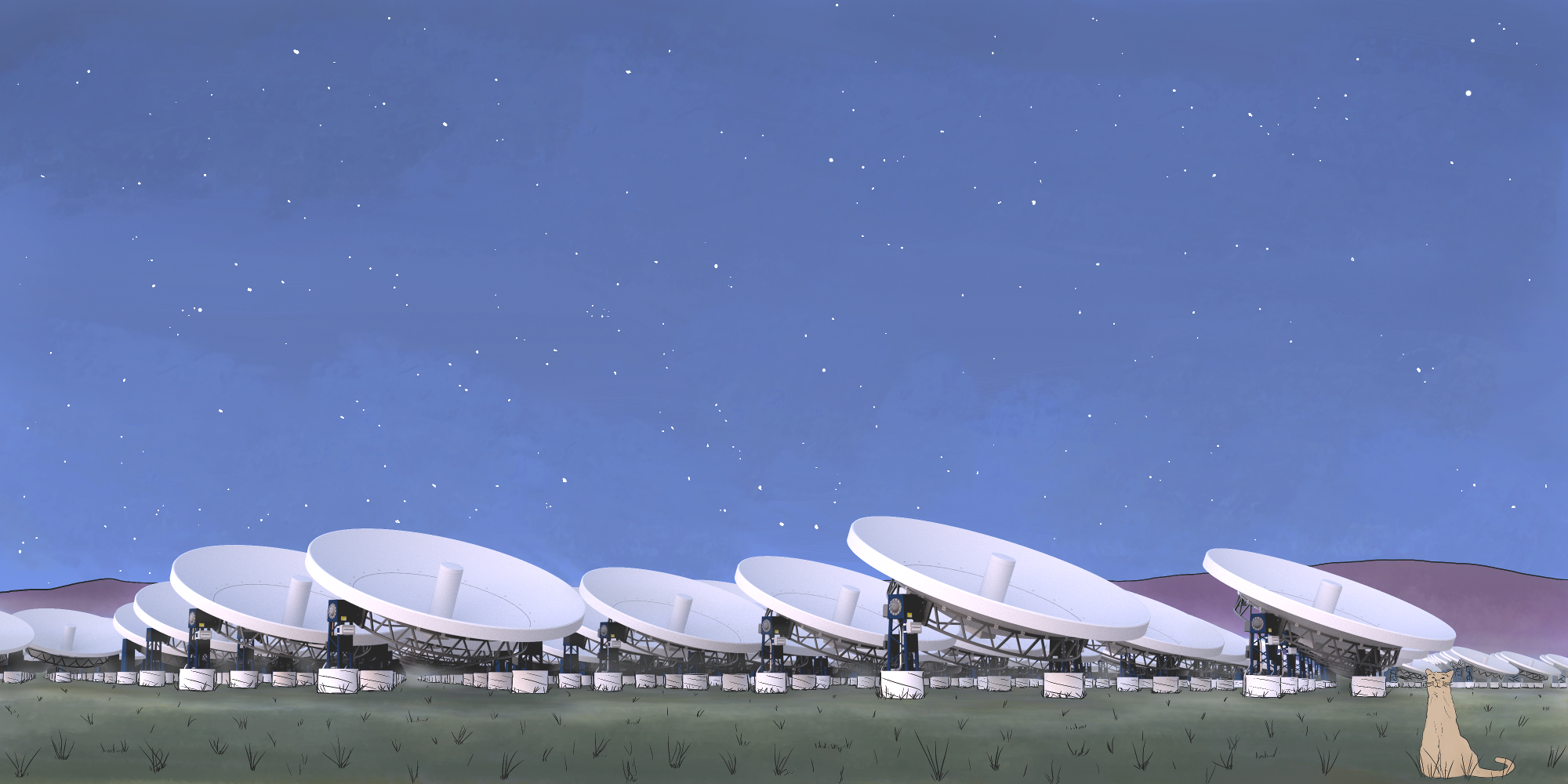}

\vspace*{0.5cm}

\textbf{Contributors and Endorsers:}
Zeeshan Ahmed$^{1}$, 
David Alonso$^{2}$, 
Mustafa A. Amin$^{3}$, 
R\'eza Ansari$^{4}$, 
Evan J. Arena$^{5,6}$, 
Kevin Bandura$^{7,8}$, 
Nicholas Battaglia$^{9}$, 
Jonathan Blazek$^{10}$, 
Philip Bull$^{11,12}$, 
Emanuele Castorina$^{13}$, 
Tzu-Ching Chang$^{14}$, 
Liam Connor$^{15}$, 
Romeel Dav\'e$^{16}$, 
Cora Dvorkin$^{17}$, 
Alexander van Engelen$^{18,19}$, 
Simone Ferraro$^{20}$, 
Raphael Flauger$^{21}$, 
Simon Foreman$^{18}$, 
Josef Frisch$^{1}$, 
Daniel Green$^{21}$, 
Gilbert Holder$^{22}$, 
Daniel Jacobs$^{19}$, 
Matthew C. Johnson$^{23,24}$, 
Joshua~S.~Dillon$^{25}$, 
Dionysios Karagiannis$^{26,27}$, 
Alexander A. Kaurov$^{28}$, 
Lloyd Knox$^{29}$, 
Adrian Liu$^{30}$, 
Marilena Loverde$^{31}$, 
Yin-Zhe Ma$^{32}$, 
Kiyoshi W. Masui$^{33}$, 
Thomas McClintock$^{5}$, 
Pieter D. Meerburg$^{34}$, 
Kavilan Moodley$^{32}$, 
Moritz M\"{u}nchmeyer$^{24}$, 
Laura B. Newburgh$^{35}$, 
Cherry Ng$^{36}$, 
Andrei Nomerotski$^{5}$, 
Paul O'Connor$^{5}$, 
Andrej Obuljen$^{37}$, 
Hamsa Padmanabhan$^{18}$, 
David Parkinson$^{38}$, 
J. Xavier Prochaska$^{39,40}$, 
Surjeet Rajendran$^{41}$, 
David Rapetti$^{42,43}$, 
Benjamin Saliwanchik$^{35}$, 
Emmanuel Schaan$^{20}$, 
Neelima Sehgal$^{44}$, 
J. Richard Shaw$^{45}$, 
Chris Sheehy$^{5}$, 
Erin Sheldon$^{5}$, 
Raphael Shirley$^{46}$, 
Eva Silverstein$^{47}$, 
Tracy Slatyer$^{33,28}$, 
An\v{z}e Slosar$^{5}$, 
Paul Stankus$^{5}$, 
Albert Stebbins$^{48}$, 
Peter T. Timbie$^{49}$, 
Gregory S. Tucker$^{50}$, 
William Tyndall$^{5,35}$, 
Francisco Villaescusa-Navarro$^{51}$, 
Benjamin Wallisch$^{28,21}$, 
Martin White$^{13,25,20}$
\vspace*{0.4cm}

{\small \noindent
$^{1}$ \SLAC \\
$^{2}$ \Oxford \\
$^{3}$ \Rice \\
$^{4}$ \ParisSud \\
$^{5}$ \BNL \\
$^{6}$ \drexel \\
$^{7}$ \WVU \\
$^{8}$ \WVUGWAC \\
$^{9}$ \Cornell \\
$^{10}$ \EPFL \\
$^{11}$ \QMUL \\
$^{12}$ \UWC \\
$^{13}$ \UCBP \\
$^{14}$ \JPL \\
$^{15}$ \AmsterdamAstro \\
$^{16}$ \ED \\
$^{17}$ \HarvardPhys \\
$^{18}$ \CITA \\
$^{19}$ \ASU \\
$^{20}$ \LBL \\
$^{21}$ \UCSD \\
$^{22}$ \UrbanaC \\
$^{23}$ \YorkU \\
$^{24}$ \PI \\
$^{25}$ \UCB \\
$^{26}$ \UNIPD \\
$^{27}$ \INFNPD \\
$^{28}$ \IAS \\
$^{29}$ \UCD \\
$^{30}$ \McGill \\
$^{31}$ \CNYang \\
$^{32}$ \KwaZuluNatal \\
$^{33}$ \MIT \\
$^{34}$ \VSI \\
$^{35}$ \Yale \\
$^{36}$ \dunlap \\
$^{37}$ \WCA \\
$^{38}$ \KASSI \\
$^{39}$ \SCIPP \\
$^{40}$ \IPMU \\
$^{41}$ \JHU \\
$^{42}$ \CUBoulder \\
$^{43}$ \NPPSFAmes \\
$^{44}$ \StonyBrook \\
$^{45}$ \UBC \\
$^{46}$ \IAC \\
$^{47}$ \Stanford \\
$^{48}$ \FNAL \\
$^{49}$ \UWMadison \\
$^{50}$ \Brown \\
$^{51}$ \CCA \\
}
}

\pagebreak
\pagenumbering{arabic}

\setlength{\parskip}{3pt}
\setlength{\parindent}{18pt}

\section{Introduction}

We present a concept for an ultra-wide band, high-sensitivity and low-resolution radio telescope array operating at \SI{200}{MHz}-\SI{1100}{MHz}, which is optimized for cosmology with \SI{21}{cm} intensity mapping in the post-reionization era, but also addresses other science goals amenable to such observations, including Fast Radio Bursts~(FRBs), pulsar monitoring and transients as a part of multi-messenger observations. Our project is named PUMA (Packed Ultra-wideband Mapping Array) and is an evolution of the \stagetwo\ \SI{21}{cm} experiment concept.

This idea grew out of deliberations of the Cosmic Visions Dark Energy Committee, a Department of Energy panel tasked with investigating possible future directions for the DOE HEP program in the field of dark energy science, and more generally low-redshift survey science. In two reports~\cite{2016arXiv160407626D,2016arXiv160407821D}, the idea of a DOE-led \SI{21}{cm} experiment was advocated, followed by the development of a concept within the \SI{21}{cm} working group. This has culminated in the \emph{21\,cm Roadmap} document~\cite{2018arXiv181009572C}, which was published on the arXiv in October 2018 and submitted to the DOE. For the decadal survey, the concept has been significantly updated following new insights, mostly regarding the feasibility of the ultra-wide bandwidth feed antennas and the theoretical modelling of the expected signal.

\section{Context}

In the next decade, three flagship US-led dark energy projects will be nearing completion: (i)~DESI, a highly multiplexed optical spectrograph on the \SI{4}{m}~Mayall telescope capable of measuring spectra of \num{5000}~objects simultaneously; (ii)~LSST, a 3~gigapixel camera on a new \SI{8}{m}-class telescope in Chile, enabling an extreme wide-field imaging survey to 27th magnitude in six filters; and (iii)~WFIRST, a space mission with a significant dark energy component, measuring both spectra and images of galaxies over very small, but very deep fields. These experiments will characterize dark energy at lower redshift with exquisite precision. Together with the continued exploration of the Cosmic Microwave Background~(CMB), they will keep the US at the forefront of cosmological observations.  Having said that, they will leave a majority of the post-reionization Universe, i.e.\ within $z<6$, uncovered -- this range can be fully surveyed via \SI{21}{cm} intensity mapping.

With PUMA, we propose a revolutionary post-DESI, post-LSST  program for dark energy, and more, based on intensity mapping of the redshifted \SI{21}{cm} emission line from neutral hydrogen; the field of observation will stretch from our local cosmological neighborhood, at $z\sim0.3$, out to $z\sim 6$, just after reionization. Unlike optical and CMB surveys, which are mature and now planning 3rd and 4th generation experiments, hydrogen intensity mapping is a relatively new technique, but one that offers important complementary science to these planned probes. The PUMA experiment has the unique capability to quadruple the volume of the Universe surveyed by optical programs, providing a percent-level measurement of the cosmic expansion history and growth to $z\sim 6$. This measurement will significantly improve the precision on standard cosmological parameters, while also opening a window for new physics beyond the concordance $\Lambda$CDM model. 

In its full configuration, the total noise will be equivalent to the sampling (Poisson) noise from a spectroscopic galaxy survey of 2.9 billion galaxies (or 600 million galaxies in the more modest, ``petite'' configuration) on large, linear scales. In addition, multiple cross-correlations with optical surveys and the CMB will dramatically improve the characterization of dark energy and new physics. The rich dataset produced by PUMA will simultaneously be useful in exploring the time-domain physics of fast radio transients and pulsars, potentially in live ``multi-messenger'' coincidence with other observatories.

PUMA is proposed with six basic science drivers in mind: two relate to fundamental advances in dark energy and modified gravity; two probe the inflationary period; and two touch on astrophysical goals, namely the detection and characterization of FRBs and pulsars. These six agendas can all be fulfilled by the same specialized instrument, as outlined in Table~\ref{tab:stm}. While these goals inform and primarily determine the design of the instrument, as with any synoptic project, PUMA will open doors to numerous other science goals which we briefly discuss in Section~\ref{sec:additional-science}.

\subsection*{Comparison with Existing Projects}

\begin{wrapfigure}[24]{r}{0.4\textwidth}
  \vspace{-15pt}
  \includegraphics[width=0.4\textwidth, trim=12 12 12 12]{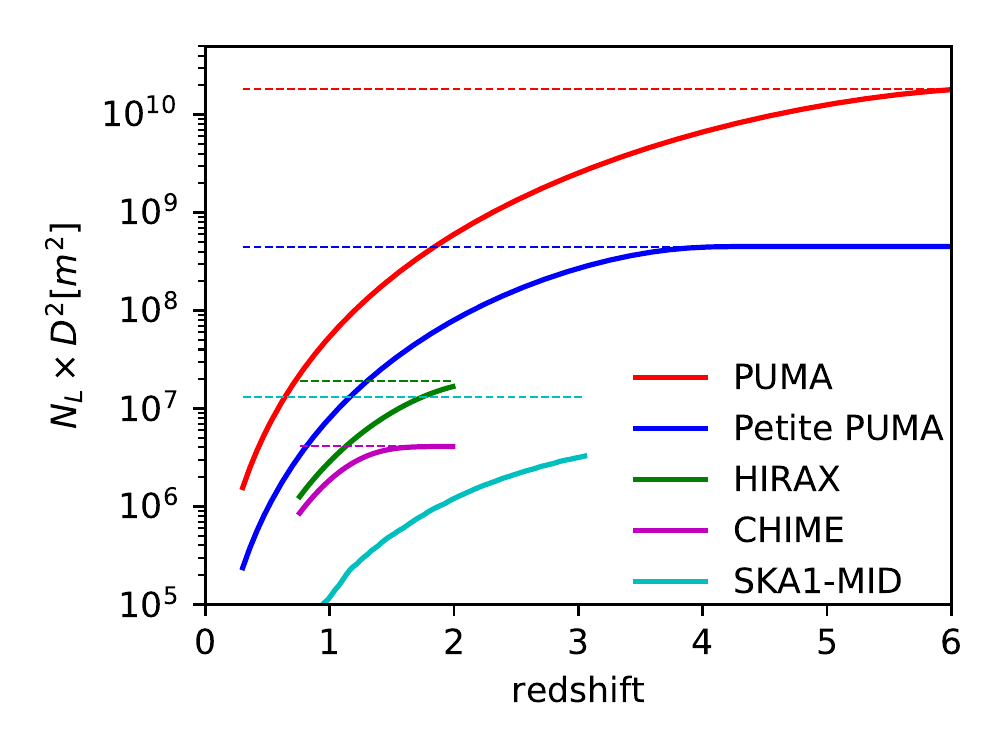}
  \captionsetup{font=footnotesize}
  \caption{The Figure of Merit (FOM) for intensity mapping experiments is the total number of baselines in the linear regime multiplied by the collecting area of each element,~$D^2$ (see endnote~\ref{ftn:D4}). We display this~FoM as a function of redshift for a selection of experiments: \stageone\ experiments (HIRAX and CHIME), the proposed \stagetwo\ experiment PUMA and a future radio telescope (SKA1-MID). Dashed lines show the value assuming all baselines in the experiment are in the linear regime. For CHIME, we take $D=({\mbox{cylinder area}}/({\mbox{feeds per cylinder}}))^{1/2}\sim \SI{2.8}{m}$. For SKA1-MID, we assume $197\times\SI{15}{m}$ dishes observing at \SI{30}{\degree} off zenith and assign a $3\times$ boost to account for better noise and optical performance. \label{fig:baselinecomp}}
\end{wrapfigure}

The design of an instrument to map large swaths of the sky at modest angular resolution, but high sensitivity, is fundamentally different from that of radio telescopes specializing in imaging of individual radio sources. This experiment is therefore not in competition with the ngVLA~\cite{1810.07524} or SKA~\cite{1311.4288}. Instead, PUMA is an evolution of a different lineage of experiments, including CHIME~\cite{CHIME} and HIRAX~\cite{HIRAX}, which we refer to as \stageone\ experiments. Compared to high-resolution imaging arrays, PUMA will be fundamentally different in three aspects:
\begin{itemize}
\item The array elements are non-tracking and with considerably simpler mechanical design, but are larger in number.
\item The array elements are closely packed together, providing information on the angular scales most important for the science outlined in Table~\ref{tab:stm}.
\item The system provides extremely large instantaneous bandwidth and relies on FFT beamforming to process the data.
\end{itemize}
In Figure~\ref{fig:baselinecomp}, we compare a few relevant experiments in terms of an intensity mapping Figure of Merit~(FOM), which is the total number of baselines probing linear scales multiplied by a single element's collecting area.\endnote{This figure of merit is motivated by the expressions for the noise power spectrum of an intensity mapping experiment, which is inversely proportional to the number of baselines, the field of view and the square of the element-collecting area (see equation~(D4) of~\cite{2018arXiv181009572C}).\label{ftn:D4}} (ngVLA is not included since it does not cover the relevant frequency range.) Despite being cheaper than SKA-MID, PUMA is the most favorable design by this metric, because it  has been optimized for this science from the very beginning. For example, at the redshifts of interest, SKA-MID will either under-resolve the baryon acoustic oscillations~(BAO) if operating in single-dish mode or over-resolve them if used as an interferometer~\cite{2017MNRAS.466.2736V}.

Of the \stageone\ experiments, the currently most advanced survey is the Canadian CHIME experiment, consisting of \num{1024}~elements and operating at 400-800\,MHz. It achieved first light in 2018 and has already published ground-breaking results on FRBs~\cite{1901.04524,1901.04525}. HIRAX is a similar experiment in the same frequency band, also with 1024 elements, but relying on dishes rather than cylindrical reflectors and is under construction in South Africa at the moment.

Many of the enabling technologies, in particular commodity DSP hardware operating at GHz frequencies developed for the telecommunications industry, ultra-wideband transducers and off-the-shelf networking of sufficient capacity, make the coming decade an ideal time for development of this project (see Section~\ref{sec:design-proj-cons}). While the FFT beamforming employed by PUMA might be one of the most ambitious aspects of this program, we remark that it is equivalent to a real-time co-addition of redundant baselines which is a technique already employed by the CHIME experiment.

\textbf{The success and lessons learned from these \stageone\ experiments will be instrumental in ensuring that PUMA delivers its science goals.}  We note that compared to these experiments, which are already operational, PUMA in its petite configuration is much more powerful, but only modestly more ambitious in terms of hardware (about five times the number of dishes and roughly twice the bandwidth). However, we plan an advanced program of technical development for system calibration and understanding of performance in advance of deploying the array.

\section{From Science Drivers to Instrument Design}

\subsection{Basic Science Drivers}

The PUMA requirements are based on three science areas with two main goals each:

\minisection{Probing the Physics of Dark Energy:}
\begin{itemize}
\item[\textbf{A.}] \textit{Characterize the expansion history in the pre-acceleration era.} By the time PUMA becomes operational, multiple experiments will have measured the expansion history close to the sample-variance limit out to redshift $z\sim 1.5$ and with some precision to $z\sim 3$. PUMA will enable nearly sample-variance limited BAO measurements all the way to $z\sim 6$ and complete the challenge of characterizing the expansion history across the cosmic ages~\cite{2019arXiv190312016S}.
  
\item[\textbf{B.}] \textit{Characterize structure growth in the pre-acceleration era.} Measuring the growth of structure over the same redshift-range as the expansion history allows fundamental tests of general relativity~\cite{2019arXiv190312016S}. If general relativity is correct, then the growth of structure is uniquely determined by the expansion history. A disagreement between the two measurements would be a smoking gun of modified gravity and PUMA is one of the most promising probes to detect it. PUMA will measure growth by relying on the weakly non-linear regime where the degeneracy between bias and growth can be broken by the shape of the power spectrum~\cite{1902.07147}.
\end{itemize}

\minisection{Probing the Physics of Inflation:}
\begin{itemize}
\item[\textbf{C.}] \textit{Constrain or detect primordial non-Gaussianity.} Primordial non-Gaussianity is one of the very few handles that we have on the physics of inflation~\cite{2019arXiv190304409M}. Departures from Gaussianity at a detectable level would provide evidence for non-minimal models of inflation which would imply either multiple fields or deviations from slow roll. It would be a monumental discovery, potentially probing physics up to the grand unification scale.
  
\item[\textbf{D.}] \textit{Constrain or detect features in the primordial power spectrum.} Features in the primordial power spectrum are another, often overlooked, handle on the physics of inflation~\cite{2019arXiv190309883S}. They are generically produced in a wide class of models of inflation and its alternatives. They could provide hints about details in the inflationary potential and, if detected, would have a profound impact on our understanding of inflation. The feasibility of this measurement in the (galaxy) power spectrum has been demonstrated in~\cite{Beutler:2019ojk}.
\end{itemize}

\minisection{Probing the Physics of the Transient Radio Sky:}
\begin{itemize}
\item[\textbf{E.}] \textit{Fast Radio Burst Tomography of the Unseen Universe.} Fast radio bursts offer a unique probe of the distant Universe if a survey with large numbers of precisely localized sources is available~\cite{2019arXiv190306535R}. Faraday rotation provides a precision probe of intergalactic magnetic fields and time-delay microlensing allows for a cosmic census of compact objects (including constraining black holes as dark matter). In addition, dispersion can be used to measure the free electron power spectrum, which breaks a degeneracy for interpretations of kinetic Sunyaev-Zeldovich~(kSZ)~measurements.
  
\item[\textbf{F.}] \textit{Monitor all pulsars discovered by SKA.} The SKA1-LOW and SKA1-MID arrays will detect of the order of \num{3000} pulsars~\cite{2015aska.confE..40K}. It is clear that none of the current telescope facilities, including SKA itself, would have enough sky time to follow up the majority of these discoveries. Due to the daily monitoring of a significant subset of these pulsars (depending on the pointing), PUMA will be complementary to SKA, even in the petite array configuration (see Figure~2F). With its unprecedentedly high timing cadence, PUMA will be able to characterize each of these new pulsar discoveries, and carry out a systematic study of pulsar temporal variabilities, including nulling, glitches, sub-pulse drifting, giant pulse emission, and potential signatures of new fundamental physics.
\end{itemize}

\noindent\textbf{From Science to Instrument.} These science drivers motivate the design of the instrument as explained in the Table~\ref{tab:stm}. In short, every intensity mapping goal translates into a natural, required resolution. This determines the longest baseline and, consequently, the linear extent of the array. The required sensitivity determines the product $N\times D$, where $N$ is the number of elements in the array and $D$ is the linear dimension of each element.\endnote{At fixed linear array size, the survey noise power spectrum scales with $N\times D \propto A_{\rm tot}/D$ since the total collecting area scales as $A_{\rm tot} \propto ND^2$.} Finally, we argue that the dishes should be as small as possible in order to allow closely-spaced baselines probing the large-scale modes that are crucial for our science goals. At the same time, in order to minimize systematics, we set $D$ to be at least a few wavelengths across at the highest redshift in order to have some primary beam localization on the sky. For now we set $D=\SI{6}{m}$ which corresponds to about four wavelengths across at the highest redshift. Science goals A to D set the basic array parameters, but the same array can also naturally achieve the science goals E and F.

\textbf{The science goals therefore naturally determine the basic parameters of the experiment,} which we list in Table~\ref{tab:array}. We note that the array parameters needed for goals A and F are considerably more relaxed compared to other science goals. However, both require a compact array filled to approximately 50\%. This leads us to a two-stage concept in which we start with a smaller array, called petite, that can achieve science goals A and F, and make inroads into all the other science goals to be followed by a full array. This two-stage approach is also attractive from the point of view of commissioning and validating the full array.

The simplified argument presented here is supported by more sophisticated forecasting, which can be found in~\cite{2018arXiv181009572C}. We conservatively account for the baseline distribution and various noise contributions, such as imperfect coupling to the sky and ground contamination. Nevertheless, some forecasts are uncertain due to our lack of knowledge of the properties of the neutral hydrogen distribution at redshifts beyond $z=2$. Similarly, the assumed rates of low-frequency FRBs are extrapolations from higher-frequencies.

\subsection{Additional Science}
\label{sec:additional-science}

The total science reach of this experiment is considerably wider than the six main goals. In the following, we provide a brief overview of some of the other exciting science capabilities of PUMA:
\begin{enumerate}[leftmargin=18pt]
\item \textbf{Broadband power spectrum information.} In addition to BAO and redshift-space distortion extraction, the exquisite precision to which the linear power spectrum can be measured will allow strong constraints on numerous basic parameters when combined with CMB and other large-scale structure data. Among the parameter constraints forecasted in~\cite{2018arXiv181009572C}, we would like to emphasize two: (i)~the error on the energy density of light relics (cf.~\cite{Green:2019glg}), parametrized by~$N_{\rm eff}$, halves to $\sigma[\Delta N_{\rm eff}]=0.013$, in combination with CMB-S4; (ii)~the dark energy equation of state, $w$, can be constrained with sub-percent precision \emph{even in cosmologies with free neutrino mass~$m_\nu$} since the $m_\nu$-$w$ degeneracy is strongly broken.

\begin{landscape}
	\begin{table}
		\scriptsize
		\begin{tabular}{p{5.1cm}|p{4.5cm}|p{5.6cm}|p{5.5cm}}
			\hline
			\textbf{Science Objective} & \textbf{Scientific Measurement Requirement} & \textbf{Measurement Objective} & \textbf{Instrument Requirements} \\
			\hline
			&   & Measure \SI{21}{cm} intensity: & \\
			\textbf{A.} Characterize expansion history                    &  Measure Baryon Acoustic Oscillations    & \quad -- over $2<z<6$ & Bandwidth must include 200-475\,MHz\\
			in the pre-acceleration era                      &  to volume-limited accuracy         & \quad -- to $k\sim 0.4 \hMpc$ & Maximum baseline $L_{\rm max}\gtrsim\SI{600}{m}$ \\
			\emph{Decadal Science Whitepaper: \cite{2019arXiv190312016S}}              &                       & \quad -- with SNR per mode $\sim 1$ at $k\sim\SI{0.2}{\hPerMpc}$ & $ND>\SI{25}{km}$ at $L_{\rm max}=\SI{600}{m}\,^{\star}$  \\
			\hline
			&                 & Measure \SI{21}{cm} intensity: & \\
			\textbf{B.} Characterize structure growth                     &  Measure growth through the \SI{21}{cm}     & \quad -- over $2<z<6$ & Bandwidth must include 200-475\,MHz\\
			in the pre-acceleration era                      &  power spectrum on weakly non-linear                 & \quad -- to $k\sim \SI{1.0}{\hPerMpc}$ & Maximum baseline $L_{\rm max}\gtrsim\SI{1500}{m}$ \\
			\emph{Decadal Science Whitepaper: \cite{2019arXiv190312016S}}                                                     & scales to volume-limited accuracy                    & \quad -- with SNR per mode $\sim 1$ at $k\sim \SI{0.6}{\hPerMpc}$ & $ND>\SI{200}{km}$ at $L_{\rm max}=\SI{1500}{m}\,^{\star}$  \\
			\hline
			
			&         Measure the \SI{21}{cm} bispectrum to  &	&                                                          \\
			\textbf{C.} Constrain or detect primordial   &          achieve non-Gaussianity sensitivity of:  &                             Measure $\gtrsim 10^9$ linear modes & Same as above plus:    \\
			non-Gaussianity  & \quad -- orthogonal: $\sigma \left(\fnl^{\rm ortho}\right) < 10$&      with SNR per mode $\sim 1$  & bandwidth $200-1100$\,MHz ($z\sim 0.3-6$) \\
			\emph{Decadal Science Whitepaper: \cite{2019arXiv190304409M}}                               &           \quad -- equiliateral: $\sigma \left(\fnl^{\rm equil}\right) < 10$ &                    &  assuming $f_{\rm sky}\sim 0.5$       \\
			\hline
			&   Measure the matter power spectrum      & & \\
			\textbf{D.} Constrain or detect features  &  over all available scales to constrain    &   Sufficient forecasted power spectrum sensitivity & Same as above \\
			in the primordial power spectrum                           &  primordial features with: &	& \\
			\emph{Decadal Science Whitepaper: \cite{2019arXiv190309883S}}                    &  \quad -- $A_{\rm lin}<\num{1e-3}$ (95\% c.l.) &	&   \\
			\hline
			& Volume limited measurement & \quad -- 1 million FRBs  &
			Fluence sensitivity threshold $ \lesssim 2.5 f_{\rm sky}^{2/3} \si{Jy.ms}$\\
			\textbf{E.} Fast Radio Burst Tomography           & of electron power spectrum, & \quad -- covering two frequency octaves &   Provide real-time FRB back-end\\ 
			\emph{Decadal Science Whitepapers: \cite{2019arXiv190306535R,2019arXiv190307370S,2019arXiv190309224K,Lynch}} & stellar mass census &\quad -- 3'' localization precision & Provide baseband buffer with triggered readout\\
			\hline
			\textbf{F.} Monitor pulsars  & Monitor all pulsars   & Detect all pulsars in current Field of View    & $10\,\sigma$ point source sensitivity $>\SI{10}{\mu Jy}$/transit\\
			\emph{Decadal Science Whitepapers: \cite{2019arXiv190308653C,2019arXiv190308194F,2019arXiv190306526L,Bower,Lynch,2019arXiv190307644K}} &
			discovered by SKA    &  brighter than \SI{10}{\mu Jy}  & Provide real-time pulsar back-end\\
			
			\hline 
			
		\end{tabular}
		
		\caption{\small Science traceability matrix\label{tab:stm} for main science drivers. All derived instrument parameters assume certain fixed system properties such as amplifier temperature, sky background and various efficiency factors as outlined in~\cite{2018arXiv181009572C}.  The total integration time is assumed to be five years. $^\star$At fixed linear dimension of the array, the noise power scales as $ND$, where $N$ is the number of elements and $D$ is their linear dimension. FRB rates and properties at frequencies below \SI{400}{MHz} are extrapolations.
		}
	\end{table}
	
	\begin{table}
		\scriptsize
		\begin{tabular}{|l|C{2.5cm}C{2.5cm}|l|}
			\hline
			Antenna Array & \multicolumn{2}{l|}{Hexagonal close-packed transit array} & \\
			& Petite & Full  &  Petite array: Achieve science goals A \& F, \\
			\quad array diameter & 600m & 1500m  &   \quad and $\sim$ 30\% of B to E  \\  
			\quad fill factor & 50\% & 50\%   &   Full array: Achieve all science goals\\
			\quad number of elements & 5,000 & 32,000  & \\
			\quad $10\,\sigma$ single transit sens. &     8.7$\mu{\rm Jy}$ & 1.3$\mu{\rm Jy}$ &  \\ 
			\hline
			Array element & \multicolumn{2}{l|}{Parabolic on-axis with N-S pointing} & transit observations, campaign repointing \\
			\quad dish diameter & \multicolumn{2}{l|}{6m} & shortest possible baselines with  $D\gg\lambda_{\rm min}$ \\
			\quad construction & \multicolumn{2}{l|}{on-site fiber-glass production, mm surface accuracy} & better beam control than \stageone\ for systematics \\
			\quad frequency coverage & \multicolumn{2}{l|}{200 -- 1100\,MHz} & \\
			\quad OMT & \multicolumn{2}{l|}{ultra-wide band, dual-pol} & \\
			\quad front-end & \multicolumn{2}{l|}{amplifiers and digitizers integrated with OMT} & alternative arrangement to be explored \\
			\quad channelizer & \multicolumn{2}{l|}{one per 10-100 dishes } & helps with corner-turning, alternatives possible \\
			\hline
			Correlator & \multicolumn{2}{l|}{FFT correlator with partial $N^2$ correlations} &  also non-FFT calibration mode\\
			\quad FRB capability & \multicolumn{2}{l|}{ real-time FRB search engine} &\\
			\quad real-time beamforming & \multicolumn{2}{l|}{$10^4$ concurrent tracking beams } & pulsar, transients, multi-messenger\\
			\hline    
		\end{tabular}
		\begin{tabular}{|l|p{5.0cm}|}
			\hline
			Survey  &   \\
			\quad area & 50\% sky \\
			\quad observing time & 5 years on sky, wall-time 7-10 years \\
			\quad equivalent source density &   \\
			\quad \quad at $z=2$, $k=\SI{0.2}{\hPerMpc}$ & $7.4 / \SI{2.0e-3}{\h\tothe3\per\Mpc\tothe3}$ (full/petite) \\
			\quad total equivalent sources  & \\
			\quad \quad at $k=\SI{0.2}{\hPerMpc}$ & 2.9/0.6  billion (full/petite)  \\ 
			\quad \quad at $k=\SI{0.5}{\hPerMpc}$ & 2.5/0.4  billion (full/petite)  \\
			\quad FRB rates (expected)  & \\
			\quad \quad  200-400\,MHz & 1200/70 per day (full/petite; \emph{uncertain})\\
			\quad \quad  400-700\,MHz & 1000/60 per day (full/petite) \\
			\quad \quad  700-1100\,MHz & 1300/80 per day (full/petite) \\
			\hline
			Calibration & \\
			\quad complex amplitude & sky sources \\
			\quad primary beam & per antenna using fixed wing drones \\
			\quad clock distribution & \SI{100}{fs} clock distribution for phase stability \\
			\hline
		\end{tabular}
		
		\caption{\label{tab:array} Basic instrumental parameters.}
	\end{table}
	
\end{landscape}

\begin{figure}[H]
	
	\hspace*{-0.6cm}
	\begin{tabular}{|c|c|c|}
		\hline
		&   {\sc \textbf{A.} Expansion History} & {\sc \textbf{B.} Growth}\\
		\begin{turn}{90}{\sc Dark Energy \&  Mod.\ Gravity}\end{turn} &    \includegraphics[width=0.47\linewidth, clip]{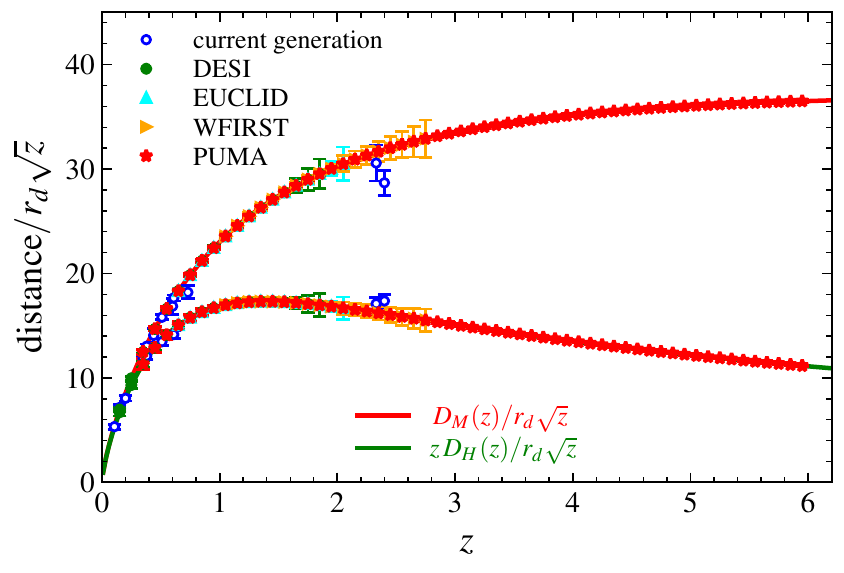} &
		\includegraphics[width=0.47\linewidth,clip]{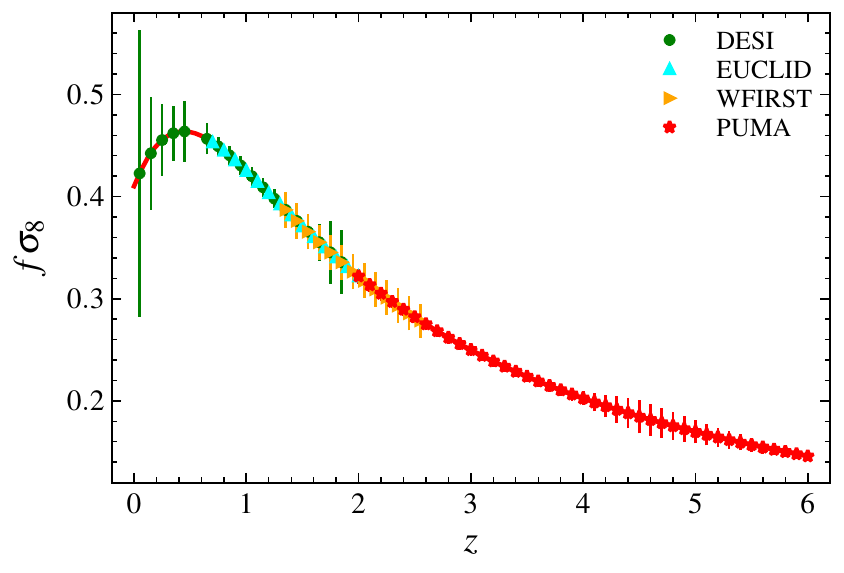} \\
		\hline
		&    {\sc \textbf{C.} Non-Gaussianity} & {\sc \textbf{D.} Inflationary Features}\\
		\begin{turn}{90} {\sc \hspace{33pt} Physics of Inflation}  \end{turn} &
		\includegraphics[width=0.48\linewidth,clip]{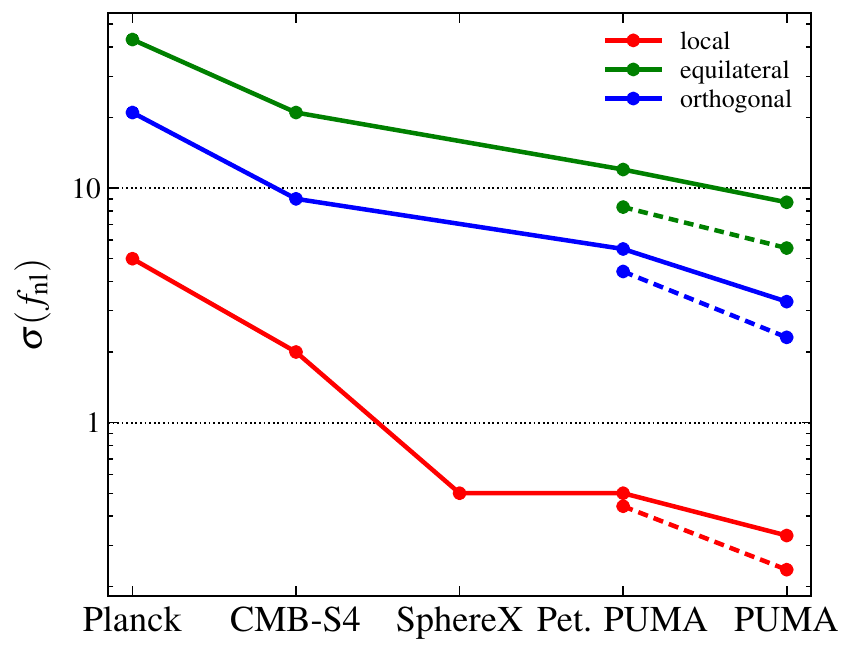}        &
		\includegraphics[width=0.48\linewidth,clip]{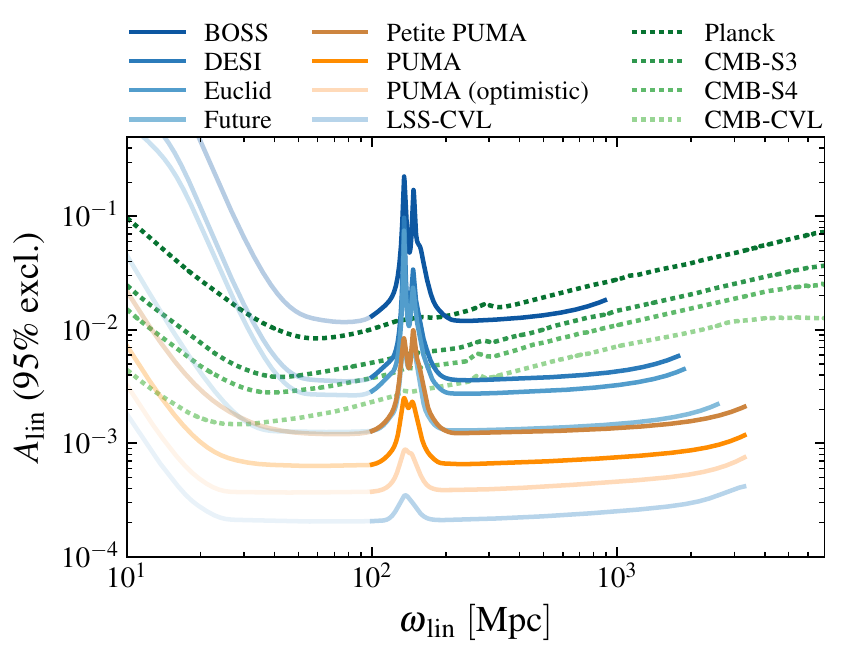}\\ 
		\hline
		&    {\sc \textbf{E.} Fast Radio Bursts} & {\sc \textbf{F.} Pulsars}\\
		\begin{turn}{+90} {\sc \hspace{10pt} Transient Radio Sky} \end{turn} &
		\includegraphics[width=0.47\linewidth,clip]{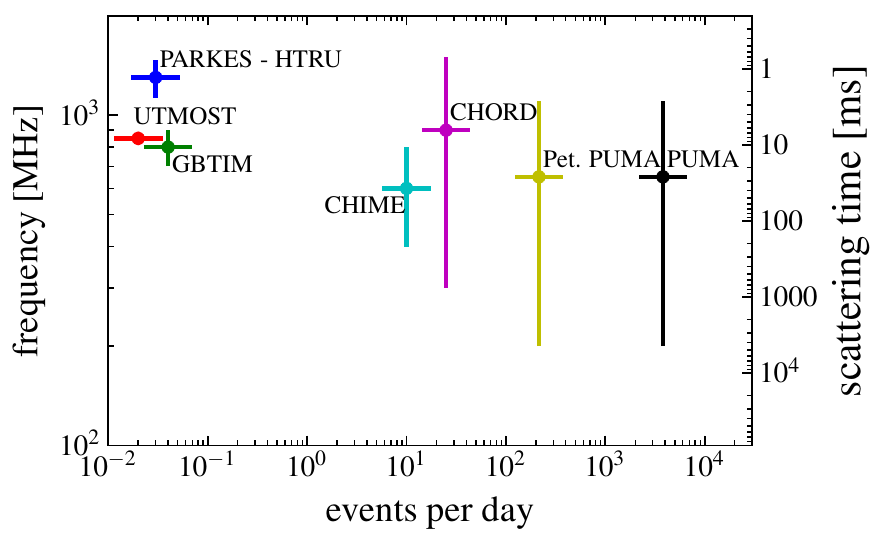}        &
		\includegraphics[width=0.47\linewidth,clip]{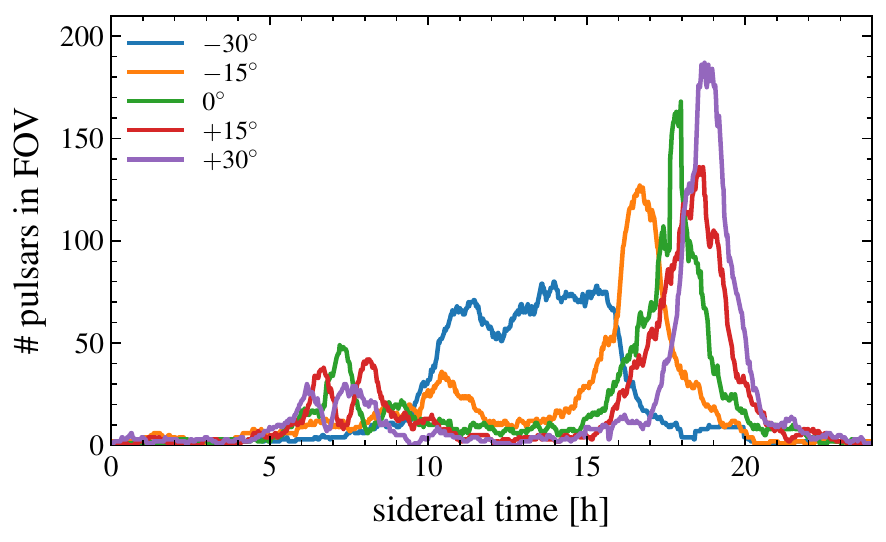}\\ 
		\hline
	\end{tabular}
	\captionsetup{font=footnotesize}
	\caption{A collage of scientific possibilities offered by PUMA (see~\cite{2018arXiv181009572C} for detailed meaning of quantities and forecasting methodology).   
			\textit{Upper Left:}~Forecasted errors of BAO-derived distances compared with current and planned experiments (\emph{petite array});
			\textit{Upper Right:}~Forecasted one-sigma errors on the growth parameter together with current and planned experiments (distinction between $\sigma_8$, $f$ and $f\sigma_8$ has been intentionally blurred for ease of comparison);
			\textit{Middle Left:}~Forecasted errors on the $\fnl$ parameters compared with current and planned experiments (dashed is for optimistic foreground control);
			\textit{Middle Right:}~Forecasted errors on linearly-spaced oscillatory features as a function of feature frequency;
			\textit{Lower Left:}~The expected daily rate of FRBs for current and future experiments;
			\textit{Lower Right:}~Known~\cite{2005AJ....129.1993M} and forecasted SKA1 pulsars in the PUMA field of view during a sidereal cycle for various N-S pointing offsets assuming observations at latitude~\SI{-30}{\degree}.}
	
\end{figure}

\item \textbf{Weak lensing \& linear matter field reconstruction.} The apparent non-Gaussianity of the observed intensity fields will be generated mostly by two mechanisms: weak lensing along the line of sight and non-linear evolution of modes. These effects can be used to reconstruct both the weak lensing potential (analogous to how the CMB measures weak lensing) and the primordial linear density field~\cite{1803.04975}. New machine-learning inspired methods have the potential to extract almost all information lost to foregrounds exploiting the non-linear cascade of power from large (contaminated) scales to smaller, well-measured scales~\cite{1907.02330}. The extraction of the primordial linear density field will allow cross-correlation with CMB lensing and galaxy lensing, but also provide a calibration of photometric redshifts in photometric surveys. Moreover, weak-lensing reconstruction will provide more lensing planes allowing for even more cross-correlations, many of which are internal to PUMA and enabled by the extremely large redshift range that is covered.

\item \textbf{Using PUMA to extract new information from the secondary CMB.}  Using Sunyaev-Zeldovich tomography, PUMA will make a tomographic reconstruction of the remote CMB dipole field~\cite{1610.06919,2018JCAP...04..034D,Deutsch:2017ybc,Cayuso:2018lhv,Smith:2018bpn} by appropriate cross-correlations with CMB observations. It can be used to reconstruct long-wavelength radial modes (e.g.\ those most affected by foregrounds in \SI{21}{cm} intensity mapping) from the statistics of small-scale transverse modes. While SZ tomography can be performed with large photometric redshift surveys such as LSST, doing this with \SI{21}{cm} intensity mapping covers a larger volume at higher signal-to-noise, potentially increasing the reach of the experiment to fundamental physics~\cite{Munchmeyer:2018eey,Cayuso:2019hen}. A particular strength of PUMA is the increased redshift range, which helps alleviate astrophysical systematic uncertainties in the forecasts by combining independent information from multiple tomographic bins~\cite{1804.10627}.

\item \textbf{Multi-messenger probes.} PUMA will have a considerable instantaneous field of view (up to 200~square degrees), which will likely cover some gravitational wave events by pure coincidence. An immediate trigger from the array of gravitational detectors expected to come online in the coming decade will allow a ring-buffer dump, enabling reconstruction of any coincident, low-frequency, electromagnetic counterpart. For transients with known locations, their fluxes could be measured by dedicating a number of real-time beamforming beams to them. The pulsar monitoring program of PUMA will also be able to associate pulsar timing glitches with bursts of gravitational waves due to crustal rearrangements~\cite{2019arXiv190309224K}.


  
\item \textbf{Time-domain survey for pulsars and magnetars.} SKA is set to be the power house in pulsar searching for the southern hemisphere. However, there is still great potential for more discoveries by re-visiting the same sky, as PUMA could observe transient phenomena such as magnetar outbursts or pulses from Rotating Radio Transients (RRATs). PUMA is designed to have a wide field-of-view, a large collecting area and a wide observing bandwidth. As we will be search for FRBs (goal~E), we will also have the capability of surveying for other time domain transients. The opportunities in the next decade to use time-domain radio astronomy to understand strong-field gravity, ultra-dense matter and high-energy astrophysics is summarized in~\cite{Lynch} (including tests of scenarios such as those laid out in~\cite{1903.04691}).
\end{enumerate}

\section{Enabling Technologies, Design and Project Considerations}
\label{sec:design-proj-cons}

There are considerable technical challenges in bringing PUMA to reality. The research roadmap is clear and outlined in related submissions to the same panel~\cite{PeterWhitePaper,AdrianWhitePaper}. We proceed assuming these challenges will be overcome, but with an implicit understanding that concrete technical solutions and, therefore, the parameters of the instrument are subject to significant changes as our understanding matures.

\pagebreak
\minisection{Site.} We assume the PUMA telescope to be based in a remote site within an established radio-quiet zone located in the southern hemisphere having a developed infrastructure and access to low-cost labor for construction and operations. The most likely candidates are existing sites for the SKA in South Africa or Australia. A southern hemisphere site will enable science objective F, while allowing cross-correlation opportunities with LSST and CMB-S4.  An alternative would be an existing NRAO site within the continental USA.

\minisection{Element construction.} We envisage reusing many of the techniques that are being currently developed for HIRAX and the proposed CHORD experiment.\endnote{CHORD is a proposed second-generation intensity mapping experiment based on the CHIME concept, planned for construction beginning in 2021 in Canada. Its present design consists of a compact array with 500 six-meter precision composite dishes instrumented with 300-1500\,MHz broadband feed and full $N^2$ correlation capability. It is complemented by several \SI{1000}{km}-baseline outrigger stations for precise sub-arcsecond transient localization. Substantial private funding has already been invested and the Canadian government will provide a funding decision in 2020. The members of the CHORD team have expressed a willingness for their demonstrated technology to be a pathfinder for more ambitious projects such as PUMA~\cite{Dobbs}.} This includes composite or molded fiberglass elements constructed on-site and receiver feeds which are supported using radio-transparent material from the dish center that provides protection from the elements as well as enabling communication wires running parallel to the ray path, minimizing diffraction. Our preliminary electromagnetic simulations show sufficient beam localization even at the lowest frequencies where the dish is only four wavelengths across.

\minisection{OMT and primary focus electronics.} Our conceptual design calls for an integrated front-end system composed of an ortho-mode transducer (OMT), amplifier and digitizer sharing the same front-end with local channelizer units serving several interferometer elements. Ultra wide-band OMTs have been in development for some time. Demonstrated dual-polarization designs have achieved over 5:1 bandwidth ratio at high coupling efficiency ($>90\%$) across the frequency band\cite{Vanderlinde}.
However, more R\&D needs to be done to achieve effective dish coupling and suitable mechanical properties for the PUMA optical design.

\minisection{Clock distribution}. In order to fully benefit from early digitization and to enable sufficient phase stability to be able to subtract foregrounds, the digitizer clocks need to be stably synchronized to within \SI{100}{fs} over hour timescales. Promising techniques to achieve such performance include synchronous enhancements to the IEEE 1588 timing protocol~\cite{Jansweijer:2013fna} and photonic phase synchronization systems such as that baselined for SKA-MID~\cite{1805.11455}. In our costing forecasts, we assume a stable clock delivered to clusters of six array elements.

\minisection{Signal transport.} Having many localized channelizer units will simplify the problem of inter-dish communication (corner-turn problem) hierarchically and the packets will be distributed to correlators using off-the-shelf networking. We will monitor the rapid anticipated technological progress during the R\&D period and implement the most cost-effective solution in the final design. The 50\% fill factor driven by the science considerations will naturally provide space for networking units connecting dishes.

\minisection{Correlator.} We will necessarily employ an FFT-based correlator that can exploit the regularity at which dishes are distributed. During calibration periods, the correlator will perform either a subset of the full $N^2$ correlation problem or use another algorithm such as EPICal~\cite{1603.02126,1904.11422}. The most cost-efficient solution at the time will be used, which will likely result in some form of GPU-based technology. With an FFT correlator, the FRB and pulsar front-ends do not need a dedicated beam-former and can instead be implemented on top of the main correlator at a modest cost. 

\minisection{Foregrounds.} Controlling foregrounds is the most challenging aspect of \SI{21}{cm} observations, because they drive the very stringent stability and dynamic range requirements. We plan a three-pronged approach: (i)~by using individual elements that go beyond what is normally required at these wavelengths in terms of surface and scattering control, (ii)~by employing a full per-element calibration scheme and (iii)~by developing new data reduction methods. All of these will require sustained R\&D in the years leading to PUMA.

\begin{wrapfigure}[9]{r}{0.30\textwidth}
  \includegraphics[width=0.3\textwidth]{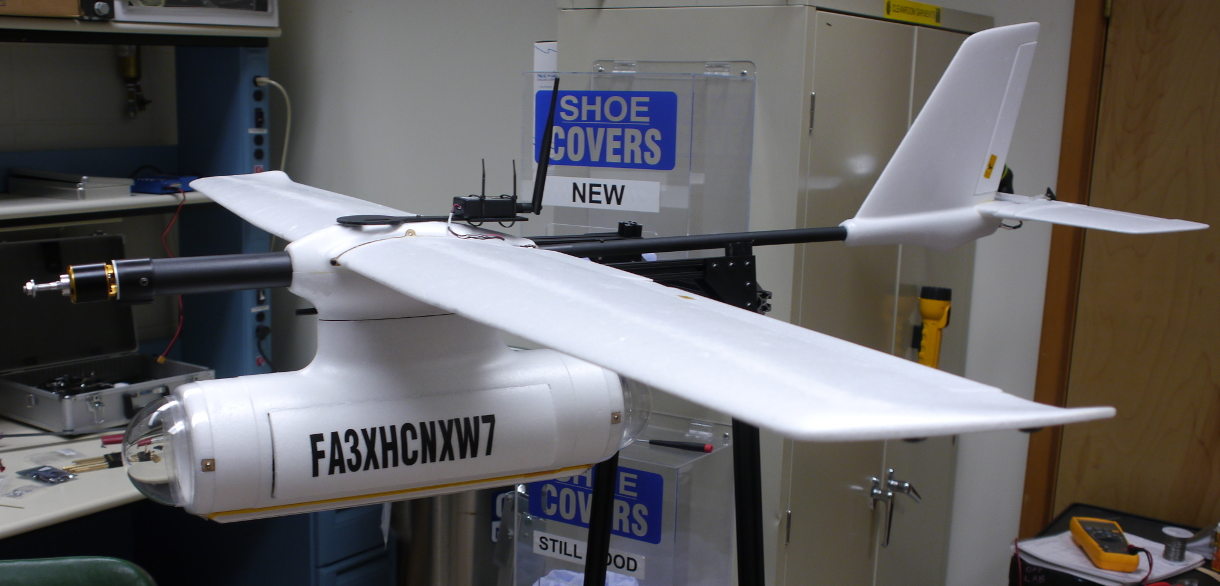}
  \captionsetup{font=footnotesize}
  \caption{Beam calibration drone in development at Brookhaven National Laboratory.}
\end{wrapfigure}
\minisection{Beam calibration and spurious coupling control.} The \stageone\ experiments are demonstrating that the knowledge of the beams of individual elements and their coupling to other nearby receivers is one of the main challenges to array performance, especially in enabling foreground separation~\cite{1810.08731}. Moreover, known and extremely stable beams are a pre-requisite for the FFT correlation required for PUMA. In addition to a tightly controlled telescope design and manufacture, we plan to dedicate substantial hardware to measuring the beam, relying on fixed-wing drones that are under development today. 

\minisection{Survey strategy.} We expect that stability of the system will be our primary concern and PUMA is therefore fundamentally a transit telescope. We plan to re-point only periodically to achieve the necessary sky coverage. The details of the survey strategy will be the result of careful simulation work as the project advances.

\minisection{Transient localization.} The current PUMA design can localize transient sources to arcminute precision which could be further improved by adding outrigger stations for a construction and operations cost increase of around 10\%.


\begin{table}[h]
  \centering
  \includegraphics[width=\linewidth,clip=true]{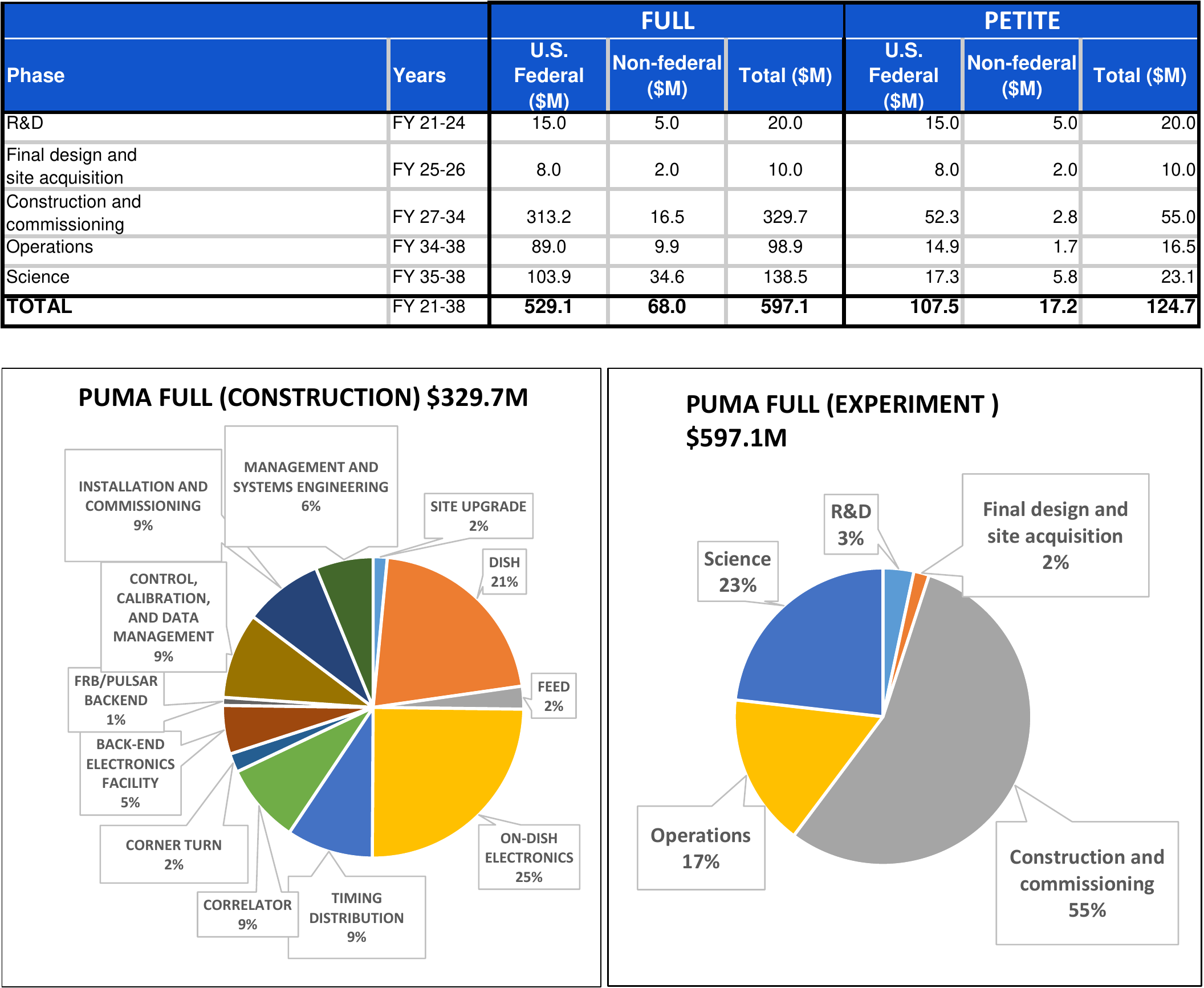}
  \caption{PUMA cost estimate. The table shows the top-level cost breakdown which is also presented in the pie-chart on the bottom right. The bottom-left pie-chart further displays the components which constitute the construction/commissioning cost. 
 }
\label{tab:costing}
\end{table}

\section{Cost Estimate}

Although the design of PUMA is at the pre-conceptual level, we will make reasonably conservative assumptions about scaling and future technology developments to extrapolate from precursor dish array projects to arrive at plausible bounds on the eventual project cost. As the design matures and the project is informed by R\&D and through experience with precursor \stageone\ surveys, a more thorough and parametric bottom-up cost estimate will be provided.

This cost estimate was performed by the Instrumentation Division at Brookhaven National Laboratory in June 2019.  The calculation was informed by input data from CHIME, HIRAX, BMX and TianLai assuming appropriate scaling and in some cases yearly cost reduction based on historical industry data. A US federal agency partnership modeled after the LSST, US-LHC and G2 dark matter experiments is anticipated to provide the bulk of the construction and operations funding, supplemented by non-federal funds from private and international participants at the 10\% level. For brevity, we only report the main numbers while the full costing details and supporting spreadsheets are available online.\endnote{\url{https://www.cosmo.bnl.gov/PUMACostingJul19.zip}}

Our operational plan is to start the experiment by building the petite array and later upgrading it to the full array. For direct comparison of costs, we however present them as two separate experiments on the same timeline. The base year is 2019 with no escalation or contingency being included in the estimate. The top level estimates for both implementation options are presented in Table~\ref{tab:costing} together with a break-down of construction costs.

\section{Conclusions}

We have presented an implementation of a \stagetwo\ \SI{21}{cm} experiment that will revolutionize the intensity mapping field and complement other planned and proposed radio experiments such as SKA, ngVLA, CHORD and DSA-2000.\endnote{DSA-2000 is a planned radio array with a focus on FRB science (see \url{http://www.astro.caltech.edu/~vikram/DSA_2000__Specifications.pdf}).} The main strength of PUMA is its sensitivity, coverage of large angular modes and very large redshift range. We have forecasted many of the science goals and presented a science-driven array design.

PUMA aims to be the first experiment to deliver an effective cost per galaxy redshift that is significantly below \$1 and, therefore, enables an effective multi-billion galaxy survey at a fraction of the cost of existing techniques.

We have outlined the expected technology evolution that will enable PUMA.  We propose that a vigorous R\&D program is started today in order to prepare and validate this concept with funding for final design and construction to follow later in this decade. The technical aspects of such a program are outlined in greater detail in~\cite{PeterWhitePaper} and many technical questions will be addressed by the current generation of experiments with similar designs (HIRAX, CHORD). The research into remaining theoretical uncertainties in the expected signal strength should be funded through standard research avenues.

\pagebreak
\section*{Acknowledgements}

The proposing team acknowledges useful feedback from volunteer experts who provided helpful comments on this text and useful discussions on the aspects of technology. These include  Matt Dobbs, Keith Vanderlinde, Rich Bradley.

We acknowledge the use of the ATNF pulsar catalogue located at \url{http://www.atnf.csiro.au/research/pulsar/psrcat/}.

\theendnotes

\bibliographystyle{utphys}
\bibliography{references}

\end{document}

%% file: institutionAliases.tex
\newcommand{\AmsterdamAstro}{Anton Pannekoek Institute for Astronomy, University of Amsterdam, 1098~XH Amsterdam, The Netherlands}

\newcommand{\ASU}{Arizona State University, Tempe, AZ~85287, USA}

\newcommand{\BNL}{Brookhaven National Laboratory, Upton, NY~11973, USA}
\newcommand{\Brown}{Brown University, Providence, RI~02912, USA}

\newcommand{\CCA}{Center for Computational Astrophysics, Flatiron Institute, New York, NY~10010, USA}

\newcommand{\CITA}{Canadian Institute for Theoretical Astrophysics, University of Toronto, Toronto, ON~M5S~3H8, Canada}

\newcommand{\CNYang}{C.N.\ Yang Institute for Theoretical Physics, State University of New York Stony Brook, NY~11794, USA}

\newcommand{\Cornell}{Cornell University, Ithaca, NY~14853, USA}

\newcommand{\CUBoulder}{Center for Astrophysics and Space Astronomy, Department of Astrophysical and Planetary Science, University of Colorado, Boulder, CO 80309, USA}

\newcommand{\drexel}{Drexel University, Philadelphia, PA~19104, USA}

\newcommand{\dunlap}{Dunlap Institute for Astronomy and Astrophysics, University of Toronto, ON~M5S~3H4, Canada}

\newcommand{\ED}{University of Edinburgh, Edinburgh EH8~9YL, UK}
\newcommand{\EPFL}{Institute of Physics, Laboratory of Astrophysics, Ecole Polytechnique Fédérale de Lausanne (EPFL), Observatoire de Sauverny, 1290 Versoix, Switzerland}

\newcommand{\FNAL}{Fermi National Accelerator Laboratory, Batavia, IL~60510, USA}

\newcommand{\HarvardPhys}{Department of Physics, Harvard University, Cambridge, MA~02138, USA}

\newcommand{\IAC}{Instituto de Astrof\'{\i}sica de Canarias, 38200 La Laguna, Tenerife, Spain}

\newcommand{\IAS}{Institute for Advanced Study, Princeton, NJ~08540, USA}

\newcommand{\INFNPD}{Istituto Nazionale di Fisica Nucleare, Sezione di Padova, 35131~Padova, Italy}

\newcommand{\IPMU}{Kavli Institute for the Physics and Mathematics of the Universe, University of Tokyo, Kashiwa, Japan}

\newcommand{\JHU}{Johns Hopkins University, Baltimore, MD~21218, USA}

\newcommand{\JPL}{Jet Propulsion Laboratory, California Institute of Technology, Pasadena, CA, USA}
\newcommand{\KASSI}{Korea Astronomy and Space Science Institute, Daejeon 34055, Korea}

\newcommand{\KwaZuluNatal}{Astrophysics and Cosmology Research Unit, School of Chemistry and Physics, University of KwaZulu-Natal, Durban 4000, South Africa}

\newcommand{\LBL}{Lawrence Berkeley National Laboratory, Berkeley, CA~94720, USA}

\newcommand{\McGill}{McGill University, Montreal, QC~H3A~2T8, Canada}

\newcommand{\MIT}{Massachusetts Institute of Technology, Cambridge, MA~02139, USA}

\newcommand{\NPPSFAmes}{NASA Postdoctoral Program Senior Fellow, NASA Ames Research Center, Moffett Field, CA 94035, USA}

\newcommand{\Oxford}{University of Oxford, Oxford OX1~3RH, UK}

\newcommand{\ParisSud}{Universit\'{e} Paris-Sud, LAL, UMR 8607, F-91898 Orsay Cedex, France \& CNRS/IN2P3, F-91405 Orsay, France}
\newcommand{\PI}{Perimeter Institute, Waterloo, ON~N2L~2Y5, Canada}

\newcommand{\QMUL}{Queen Mary University of London, London E1~4NS, UK}

\newcommand{\Rice}{Department of Physics \& Astronomy, Rice University, Houston, Texas 77005, USA}

\newcommand{\SCIPP}{University of California at Santa Cruz, Santa Cruz, CA~95064, USA}

\newcommand{\SLAC}{SLAC National Accelerator Laboratory, Menlo Park, CA~94025, USA}

\newcommand{\Stanford}{Stanford University, Stanford, CA~94305, USA}
\newcommand{\StonyBrook}{Stony Brook University, Stony Brook, NY~11794, USA}

\newcommand{\UBC}{University of British Columbia, Vancouver, BC V6T 1Z1, Canada}
\newcommand{\UCB}{Department of Astronomy, University of California Berkeley, Berkeley, CA~94720, USA}
\newcommand{\UCBP}{Department of Physics, University of California Berkeley, Berkeley, CA~94720, USA}

\newcommand{\UCD}{University of California at Davis, Davis, CA~95616, USA}

\newcommand{\UCSD}{University of California San Diego, La Jolla, CA~92093, USA}

\newcommand{\UNIPD}{Dipartimento di Fisica e Astronomia ``G. Galilei'', Universit\`a degli Studi di Padova, 35131~Padova, Italy}

\newcommand{\UrbanaC}{Department of Physics, University of Illinois at Urbana-Champaign, Urbana, IL~61801, USA}

\newcommand{\UWMadison}{Department of Physics, University of Wisconsin-Madison, Madison, WI~53706, USA}

\newcommand{\UWC}{Department of Physics \& Astronomy, University of the Western Cape, Cape Town 7535, South Africa}

\newcommand{\VSI}{Van Swinderen Institute for Particle Physics and Gravity, University of Groningen, 9747~AG~Groningen, The~Netherlands}

\newcommand{\WCA}{Centre for Astrophysics, University of Waterloo, Waterloo, ON~N2L~3G1, Canada}

\newcommand{\WVU}{CSEE, West Virginia University, Morgantown, WV 26505, USA}
\newcommand{\WVUGWAC}{Center for Gravitational Waves and Cosmology, West Virginia University, Morgantown, WV 26505, USA}

\newcommand{\Yale}{Department of Physics, Yale University, New Haven, CT~06520, USA}
\newcommand{\YorkU}{Department of Physics and Astronomy, York University, Toronto, ON~M3J~1P3, Canada}

%% file: main.bbl
\providecommand{\href}[2]{#2}\begingroup\raggedright\begin{thebibliography}{10}

\bibitem{2016arXiv160407626D}
S.~Dodelson, K.~Heitmann, C.~Hirata, K.~Honscheid, A.~Roodman, U.~Seljak,
  A.~Slosar, and M.~Trodden, ``{Cosmic Visions Dark Energy: Science},''
\href{http://arxiv.org/abs/1604.07626}{{\ttfamily arXiv:1604.07626
  [astro-ph.CO]}}.

\bibitem{2016arXiv160407821D}
S.~Dodelson, K.~Heitmann, C.~Hirata, K.~Honscheid, A.~Roodman, U.~Seljak,
  A.~Slosar, and M.~Trodden, ``{Cosmic Visions Dark Energy: Technology},''
\href{http://arxiv.org/abs/1604.07821}{{\ttfamily arXiv:1604.07821
  [astro-ph.IM]}}.

\bibitem{2018arXiv181009572C}
{R. Ansari \textit{et al.} (Cosmic Visions \SI{21}{cm} Collaboration)},
  ``{Inflation and Early Dark Energy with a Stage II Hydrogen Intensity Mapping
  experiment},''
\href{http://arxiv.org/abs/1810.09572}{{\ttfamily arXiv:1810.09572
  [astro-ph.CO]}}.

\bibitem{1810.07524}
E.~J. Murphy {\em et~al.}, ``{Science with an ngVLA: The ngVLA Science Case and
  Associated Science Requirements},'' {\em ASP Conf. Ser.} {\bfseries 517}
  (2018) 3,
\href{http://arxiv.org/abs/1810.07524}{{\ttfamily arXiv:1810.07524
  [astro-ph.IM]}}.

\bibitem{1311.4288}
M.~Huynh and J.~Lazio, ``{An Overview of the Square Kilometre Array},''
\href{http://arxiv.org/abs/1311.4288}{{\ttfamily arXiv:1311.4288
  [astro-ph.IM]}}.

\bibitem{CHIME}
K.~Bandura {\em et~al.}, ``{Canadian Hydrogen Intensity Mapping Experiment
  (CHIME) Pathfinder},'' \href{http://dx.doi.org/10.1117/12.2054950}{{\em Proc.
  SPIE Int. Soc. Opt. Eng.} {\bfseries 9145} (2014) 22},
\href{http://arxiv.org/abs/1406.2288}{{\ttfamily arXiv:1406.2288
  [astro-ph.IM]}}.

\bibitem{HIRAX}
L.~B. Newburgh {\em et~al.}, ``{HIRAX: A Probe of Dark Energy and Radio
  Transients},'' \href{http://dx.doi.org/10.1117/12.2234286}{{\em Proc. SPIE
  Int. Soc. Opt. Eng.} {\bfseries 9906} (2016) 99065X},
\href{http://arxiv.org/abs/1607.02059}{{\ttfamily arXiv:1607.02059
  [astro-ph.IM]}}.

\bibitem{2017MNRAS.466.2736V}
F.~Villaescusa-Navarro, D.~Alonso, and M.~Viel, ``{Baryonic Acoustic
  Oscillations from \SI{21}{cm} Intensity Mapping: The Square Kilometre Array
  Case},'' \href{http://dx.doi.org/10.1093/mnras/stw3224}{{\em MNRAS}
  {\bfseries 466} (2017) 2736},
\href{http://arxiv.org/abs/1609.00019}{{\ttfamily arXiv:1609.00019
  [astro-ph.CO]}}.

\bibitem{1901.04524}
{M. Amiri \textit{et al.} (CHIME/FRB Collaboration)}, ``{Observations of Fast
  Radio Bursts at Frequencies Down to 400 Megahertz},''
  \href{http://dx.doi.org/10.1038/s41586-018-0867-7}{{\em Nature} {\bfseries
  566} (2019) 230},
\href{http://arxiv.org/abs/1901.04524}{{\ttfamily arXiv:1901.04524
  [astro-ph.HE]}}.

\bibitem{1901.04525}
{M. Amiri \textit{et al.} (CHIME/FRB Collaboration)}, ``{A Second Source of
  Repeating Fast Radio Bursts},''
  \href{http://dx.doi.org/10.1038/s41586-018-0864-x}{{\em Nature} {\bfseries
  566} (2019) 235},
\href{http://arxiv.org/abs/1901.04525}{{\ttfamily arXiv:1901.04525
  [astro-ph.HE]}}.

\bibitem{2019arXiv190312016S}
A.~Slosar, T.~Davis, D.~Eisenstein, R.~Hložek, M.~Ishak-Boushaki,
  R.~Mandelbaum, P.~Marshall, J.~Sakstein, and M.~White, ``{Dark Energy and
  Modified Gravity},'' {\em \baas} {\bfseries 51} (2019) 97,
\href{http://arxiv.org/abs/1903.12016}{{\ttfamily arXiv:1903.12016
  [astro-ph.CO]}}.

\bibitem{1902.07147}
E.~Castorina and M.~White, ``{Measuring the Growth of Structure with Intensity
  Mapping Surveys},''
  \href{http://dx.doi.org/10.1088/1475-7516/2019/06/025}{{\em JCAP} {\bfseries
  06} (2019) 025},
\href{http://arxiv.org/abs/1902.07147}{{\ttfamily arXiv:1902.07147
  [astro-ph.CO]}}.

\bibitem{2019arXiv190304409M}
P.~D. Meerburg {\em et~al.}, ``{Primordial Non-Gaussianity},'' {\em \baas}
  {\bfseries 51} (2019) 107,
\href{http://arxiv.org/abs/1903.04409}{{\ttfamily arXiv:1903.04409
  [astro-ph.CO]}}.

\bibitem{2019arXiv190309883S}
A.~{Slosar}, X.~{Chen}, C.~{Dvorkin}, D.~{Green}, D.~{Meerburg},
  E.~{Silverstein}, and B.~{Wallisch}, ``{Scratches from the Past: Inflationary
  Archaeology through Features in the Power Spectrum of Primordial
  Fluctuations},'' {\em \baas} {\bfseries 51} (2019) 98,
\href{http://arxiv.org/abs/1903.09883}{{\ttfamily arXiv:1903.09883
  [astro-ph.CO]}}.

\bibitem{Beutler:2019ojk}
F.~Beutler, M.~Biagetti, D.~Green, A.~Slosar, and B.~Wallisch, ``{Primordial
  Features from Linear to Nonlinear Scales},''
\href{http://arxiv.org/abs/1906.08758}{{\ttfamily arXiv:1906.08758
  [astro-ph.CO]}}.

\bibitem{2019arXiv190306535R}
V.~Ravi {\em et~al.}, ``{Fast Radio Burst Tomography of the Unseen Universe},''
  {\em \baas} {\bfseries 51} (2019) 420,
\href{http://arxiv.org/abs/1903.06535}{{\ttfamily arXiv:1903.06535
  [astro-ph.HE]}}.

\bibitem{2015aska.confE..40K}
E.~F. Keane {\em et~al.}, ``{A Cosmic Census of Radio Pulsars with the SKA},''
  \href{http://dx.doi.org/10.22323/1.215.0040}{{\em PoS} {\bfseries AASKA14}
  (2015) 040},
\href{http://arxiv.org/abs/1501.00056}{{\ttfamily arXiv:1501.00056
  [astro-ph.IM]}}.

\bibitem{Green:2019glg}
D.~{Green}, M.~A. {Amin}, J.~{Meyers}, and B.~{Wallisch}, ``{Messengers from
  the Early Universe: Cosmic Neutrinos and Other Light Relics},'' {\em \baas}
  {\bfseries 51} (2019) 159,
\href{http://arxiv.org/abs/1903.04763}{{\ttfamily arXiv:1903.04763
  [astro-ph.CO]}}.

\bibitem{2019arXiv190307370S}
D.~R. {Stinebring}, S.~{Chatterjee}, S.~E. {Clark}, J.~M. {Cordes}, T.~{Dolch},
  C.~{Heiles}, A.~S. {Hill}, M.~{Jones}, V.~{Kaspi}, and M.~T. {Lam}, ``{Twelve
  Decades: Probing the Interstellar Medium from Kiloparsec to Sub-AU Scales},''
  {\em \baas} {\bfseries 51} (2019) 492,
  \href{http://arxiv.org/abs/1903.07370}{{\ttfamily arXiv:1903.07370
  [astro-ph.GA]}}.

\bibitem{2019arXiv190309224K}
V.~Kalogera {\em et~al.}, ``{The Yet-Unobserved Multi-Messenger
  Gravitational-Wave Universe},'' {\em \baas} {\bfseries 51} (2019) 239,
\href{http://arxiv.org/abs/1903.09224}{{\ttfamily arXiv:1903.09224
  [astro-ph.HE]}}.

\bibitem{Lynch}
R.~{Lynch}, P.~{Brook}, S.~{Chatterjee}, T.~{Dolch}, M.~{Kramer}, M.~T. {Lam},
  N.~{Lewandowska}, M.~{McLaughlin}, N.~{Pol}, and I.~{Stairs}, ``{The Virtues
  of Time and Cadence for Pulsars and Fast Transients},'' {\em \baas}
  {\bfseries 51} (2019) 461.

\bibitem{2019arXiv190308653C}
{J. M. Cordes and M. A. McLaughlin (for the NANOGrav Collaboration)},
  ``{Gravitational Waves, Extreme Astrophysics and Fundamental Physics with
  Precision Pulsar Timing},'' {\em \baas} {\bfseries 51} (2019) 447,
\href{http://arxiv.org/abs/1903.08653}{{\ttfamily arXiv:1903.08653
  [astro-ph.IM]}}.

\bibitem{2019arXiv190308194F}
E.~Fonseca, P.~Demorest, S.~Ransom, and I.~Stairs, ``{Fundamental Physics with
  Radio Millisecond Pulsars},'' {\em \baas} {\bfseries 51} (2019) 425,
\href{http://arxiv.org/abs/1903.08194}{{\ttfamily arXiv:1903.08194
  [astro-ph.HE]}}.

\bibitem{2019arXiv190306526L}
D.~R. Lorimer {\em et~al.}, ``{Radio Pulsar Populations},'' {\em \baas}
  {\bfseries 51} (2019) 261,
\href{http://arxiv.org/abs/1903.06526}{{\ttfamily arXiv:1903.06526
  [astro-ph.HE]}}.

\bibitem{Bower}
G.~{Bower}, S.~{Chatterjee}, J.~{Cordes}, P.~{Demorest}, J.~S. {Deneva},
  J.~{Dexter}, R.~{Eatough}, M.~{Kramer}, J.~{Lazio}, K.~{Liu}, S.~Ransom,
  L.~Shao, N.~Wex, and R.~Wharton, ``{Fundamental Physics with Galactic Center
  Pulsars},'' {\em \baas} {\bfseries 51} (2019) 438.

\bibitem{2019arXiv190307644K}
{L. Z. Kelley \textit{et al.} (NANOGrav Collaboration)}, ``{Multi-Messenger
  Astrophysics with Pulsar Timing Arrays},'' {\em \baas} {\bfseries 51} (2019)
  490,
\href{http://arxiv.org/abs/1903.07644}{{\ttfamily arXiv:1903.07644
  [astro-ph.HE]}}.

\bibitem{2005AJ....129.1993M}
R.~N. Manchester, G.~B. Hobbs, A.~Teoh, and M.~Hobbs, ``{The Australia
  Telescope National Facility Pulsar Catalogue},''
  \href{http://dx.doi.org/10.1086/428488}{{\em Astron. J.} {\bfseries 129}
  (2005) 1993},
\href{http://arxiv.org/abs/astro-ph/0412641}{{\ttfamily arXiv:astro-ph/0412641
  [astro-ph]}}.

\bibitem{1803.04975}
S.~Foreman, P.~D. Meerburg, A.~van Engelen, and J.~Meyers, ``{Lensing
  Reconstruction from Line Intensity Maps: The Impact of Gravitational
  Nonlinearity},'' \href{http://dx.doi.org/10.1088/1475-7516/2018/07/046}{{\em
  JCAP} {\bfseries 07} (2018) 046},
\href{http://arxiv.org/abs/1803.04975}{{\ttfamily arXiv:1803.04975
  [astro-ph.CO]}}.

\bibitem{1907.02330}
C.~Modi, M.~White, A.~Slosar, and E.~Castorina, ``{Reconstructing Large-Scale
  Structure with Neutral Hydrogen Surveys},''
\href{http://arxiv.org/abs/1907.02330}{{\ttfamily arXiv:1907.02330
  [astro-ph.CO]}}.

\bibitem{1610.06919}
A.~Terrana, M.-J. Harris, and M.~C. Johnson, ``{Analyzing the Cosmic Variance
  Limit of Remote Dipole Measurements of the Cosmic Microwave Background using
  the Large-Scale Kinetic Sunyaev-Zel'dovich Effect},''
  \href{http://dx.doi.org/10.1088/1475-7516/2017/02/040}{{\em JCAP} {\bfseries
  02} (2017) 040},
\href{http://arxiv.org/abs/1610.06919}{{\ttfamily arXiv:1610.06919
  [astro-ph.CO]}}.

\bibitem{2018JCAP...04..034D}
A.-S. Deutsch, M.~C. Johnson, M.~Münchmeyer, and A.~Terrana, ``{Polarized
  Sunyaev-Zel'dovich Tomography},''
  \href{http://dx.doi.org/10.1088/1475-7516/2018/04/034}{{\em JCAP} {\bfseries
  04} (2018) 034},
\href{http://arxiv.org/abs/1705.08907}{{\ttfamily arXiv:1705.08907
  [astro-ph.CO]}}.

\bibitem{Deutsch:2017ybc}
A.-S. Deutsch, E.~Dimastrogiovanni, M.~C. Johnson, M.~Münchmeyer, and
  A.~Terrana, ``{Reconstruction of the Remote Dipole and Quadrupole Fields from
  the Kinetic Sunyaev-Zel’dovich and Polarized Sunyaev-Zel’dovich
  Effects},'' \href{http://dx.doi.org/10.1103/PhysRevD.98.123501}{{\em Phys.
  Rev. D} {\bfseries 98} (2018) 123501},
\href{http://arxiv.org/abs/1707.08129}{{\ttfamily arXiv:1707.08129
  [astro-ph.CO]}}.

\bibitem{Cayuso:2018lhv}
J.~I. Cayuso, M.~C. Johnson, and J.~B. Mertens, ``{Simulated Reconstruction of
  the Remote Dipole Field Using the Kinetic Sunyaev-Zel’dovich effect},''
  \href{http://dx.doi.org/10.1103/PhysRevD.98.063502}{{\em Phys. Rev. D}
  {\bfseries 98} (2018) 063502},
\href{http://arxiv.org/abs/1806.01290}{{\ttfamily arXiv:1806.01290
  [astro-ph.CO]}}.

\bibitem{Smith:2018bpn}
K.~M. Smith, M.~S. Madhavacheril, M.~Münchmeyer, S.~Ferraro, U.~Giri, and
  M.~C. Johnson, ``{KSZ Tomography and the Bispectrum},''
\href{http://arxiv.org/abs/1810.13423}{{\ttfamily arXiv:1810.13423
  [astro-ph.CO]}}.

\bibitem{Munchmeyer:2018eey}
M.~Münchmeyer, M.~S. Madhavacheril, S.~Ferraro, M.~C. Johnson, and K.~M.
  Smith, ``{Constraining Local Non-Gaussianities with kSZ Tomography},''
\href{http://arxiv.org/abs/1810.13424}{{\ttfamily arXiv:1810.13424
  [astro-ph.CO]}}.

\bibitem{Cayuso:2019hen}
J.~I. Cayuso and M.~C. Johnson, ``{Towards Testing CMB Anomalies using the
  Kinetic and Polarized Sunyaev-Zel'dovich Effects},''
\href{http://arxiv.org/abs/1904.10981}{{\ttfamily arXiv:1904.10981
  [astro-ph.CO]}}.

\bibitem{1804.10627}
H.~Padmanabhan, A.~Refregier, and A.~Amara, ``{Impact of Astrophysics on
  Cosmology Forecasts for \SI{21}{cm} Surveys},''
  \href{http://dx.doi.org/10.1093/mnras/stz683}{{\em MNRAS} {\bfseries 485}
  (2019) 4060},
\href{http://arxiv.org/abs/1804.10627}{{\ttfamily arXiv:1804.10627
  [astro-ph.CO]}}.

\bibitem{1903.04691}
C.~Law {\em et~al.}, ``{Radio Time-Domain Signatures of Magnetar Birth},'' {\em
  \baas} {\bfseries 51} (2019) 319,
\href{http://arxiv.org/abs/1903.04691}{{\ttfamily arXiv:1903.04691
  [astro-ph.HE]}}.

\bibitem{PeterWhitePaper}
P.~Timbie {\em et~al.}, ``{White Paper on R\&D for HI Intensity Mapping},''
  {\em Submission to Astrophysics and Astronomy Decadal Survey} (July 2019) .

\bibitem{AdrianWhitePaper}
{The Hydrogen Epoch of Reionization Array (HERA) Collaboration}, ``{A Roadmap
  for Astrophysics and Cosmology with High-Redshift \SI{21}{cm} Intensity
  Mapping},'' {\em Submission to Astrophysics and Astronomy Decadal Survey}
  (July 2019) .

\bibitem{Vanderlinde}
K.~Vanderlinde. {Private Communication}, 2019.

\bibitem{Jansweijer:2013fna}
P.~P.~M. Jansweijer, H.~Z. Peek, and E.~de~Wolf, ``{White Rabbit:
  Sub-Nanosecond Timing over Ethernet},''
\href{http://dx.doi.org/10.1016/j.nima.2012.12.096}{{\em Nucl. Instrum. Meth.
  A} {\bfseries 725} (2013) 187}.

\bibitem{1805.11455}
S.~W. {Schediwy}, D.~R. {Gozzard}, C.~{Gravestock}, S.~{Stobie}, R.~{Whitaker},
  J.~A. {Malan}, P.~{Boven}, and K.~{Grainge}, ``{The Mid-Frequency Square
  Kilometre Array Phase Synchronisation System},''
  \href{http://dx.doi.org/10.1017/pasa.2018.48}{{\em \pasa} {\bfseries 36}
  (2019) e007}, \href{http://arxiv.org/abs/1805.11455}{{\ttfamily
  arXiv:1805.11455 [astro-ph.IM]}}.

\bibitem{1603.02126}
A.~P. {Beardsley}, N.~{Thyagarajan}, J.~D. {Bowman}, and M.~F. {Morales}, ``{An
  Efficient Feedback Calibration Algorithm for Direct Imaging Radio
  Telescopes},'' \href{http://dx.doi.org/10.1093/mnras/stx1512}{{\em MNRAS}
  {\bfseries 470} (2017) 4720},
  \href{http://arxiv.org/abs/1603.02126}{{\ttfamily arXiv:1603.02126
  [astro-ph.IM]}}.

\bibitem{1904.11422}
J.~{Kent}, J.~{Dowell}, A.~{Beardsley}, N.~{Thyagarajan}, G.~{Taylor}, and
  J.~{Bowman}, ``{A Real-Time, All-Sky, High Time Resolution, Direct Imager for
  the Long-Wavelength Array},''
  \href{http://dx.doi.org/10.1093/mnras/stz1206}{{\em MNRAS} {\bfseries 486}
  (2019) 5052}, \href{http://arxiv.org/abs/1904.11422}{{\ttfamily
  arXiv:1904.11422 [astro-ph.IM]}}.

\bibitem{1810.08731}
M.~F. Morales, A.~Beardsley, J.~Pober, N.~Barry, B.~Hazelton, D.~Jacobs, and
  I.~Sullivan, ``{Understanding the Diversity of \SI{21}{cm} Cosmology
  Analyses},'' \href{http://dx.doi.org/10.1093/mnras/sty2844}{{\em MNRAS}
  {\bfseries 483} (2019) 2207},
\href{http://arxiv.org/abs/1810.08731}{{\ttfamily arXiv:1810.08731
  [astro-ph.CO]}}.

\bibitem{Dobbs}
M.~Dobbs. {Private Communication}, 2019.

\end{thebibliography}\endgroup
